\documentclass[aps,pra,reprint,groupedaddress,showpacs]{revtex4-1}

\usepackage{amsmath}
\usepackage{graphicx}
\usepackage{textcomp}
\usepackage{hyperref}

\begin{document}

\title{Reexamination of group velocities of structured light pulses}

\author{Peeter Saari}
\email[]{peeter.saari@ut.ee}

\affiliation{Institute of Physics, University of Tartu, W. Ostwaldi 1, 50411, Tartu, Estonia}
\affiliation{Estonian Academy of Sciences, Kohtu 6, 10130 Tallinn, Estonia}

\date{\today}

\begin{abstract}
Recently a series of theoretical and experimental papers on free-space 
propagation of pulsed Laguerre-Gaussian and Bessel beams was published, which 
reached contradictory and controversial results about group velocities of such 
pulses.
Depending on the measurement scheme, the group velocity can be defined 
differently.
We analyze how different versions of group velocity are related to the 
measurable travel time (time of flight) of the pulse between input (source)  and 
output (detecting) planes.
The analysis is tested on a theoretical model---the Bessel-Gauss pulse whose 
propagation path exhibits both subluminal and superluminal regions.
Our main conclusion from resolving the contradictions in the literature is that 
different versions of group velocity are appropriate, depending on whether or not the 
beam is hollow and how the pulse is recorded in the output 
plane---integrally or with spatial resolution.
\end{abstract}

\maketitle

\section{Introduction}

Structured light fields, particularly the non-diffracting and twisted beams are 
being increasingly used in different fields of research (see reviews 
\cite{roadmap,OAM,LW2}).
Recently a series of papers \cite{Gio,Horv2015,Bareza,Bouch,MinuComm1,MajoranaTorn,Alf,MinuComm2,Faccio2,Sirtsuga}
was published, which deal with slower than $c$ group velocities of pulses of
these beams in free space and indicate promising applications of such
subluminal light propagation.

In Ref. \cite{Gio}, using time-correlated photon pairs and a 
sophisticated measurement of propagation delays \textit{via} the Hong-Ou-Mandel 
dip, the subluminality of photons in both a Bessel beam and in a focused 
Gaussian beam was shown.
This paper garnered substantial coverage in the general media because the study 
purported to discover subluminal photons.
Since at least for the physical optics community the observed phenomenon had 
been well-known,  the paper \cite{Gio} encountered significant criticism.
In particular, a comment \cite{Horv2015} finds the interpretation of the 
results and the title of the paper misleading and states that the measurements 
only provide the projection of the photon velocity onto the axis of beam 
propagation.
From our point of view, the most valuable contribution of Ref.
\cite{Gio} is a new concept of spatially averaged group velocity and its 
theoretical reasoning.
As we will see below, although this new definition of group velocity 
corresponds well to a particular type of time-of-flight measurements of light 
pulses, it leads to a contradiction in the case of certain Bessel beam pulses.

The next theoretical paper \cite{Bareza} generalizes the concept of spatially 
averaged group velocity to Gaussian beams with orbital angular momentum (OAM).
The same (twisted) beams were studied in Ref.
\cite{Bouch}, where curves of group velocity, calculated from the common (not 
spatially averaged) expression, were related to experimental data.
However, our analysis \cite{MinuComm1}  led to the conclusion that the results 
of \cite{Bouch} are questionable in several respects and, in particular, that 
when interpreting the measured relatively large propagation delays one must not 
use the absolute value of the group velocity as done in  \cite{Bouch}, but, 
instead, its projections onto the beam axis.
The very recent paper \cite{MajoranaTorn} shows that the group velocity obeys a 
relationship similar to one proposed 80 years ago by Majorana between spin and 
mass for relativistic particles.
This paper also points out that what was measured in Refs.
\cite{Gio,Bouch} as group velocity was in fact its projection onto the 
beam propagation axis.

In the theoretical paper \cite{Alf} reduction of the group velocity (not 
spatially averaged) below the value $c$ in the case of certain Bessel beam 
pulses has been considered.
We pointed out in our critical comment \cite{MinuComm2} that the authors treat 
the problem as if only one type of Bessel pulse exists, no matter how it is 
generated, while it is well known from the literature that such pulses may be not 
only subluminal but superluminal as well.

The very recent study \cite{Faccio2} which is a continuation of the work 
\cite{Gio}  and uses the same experimental technique, deals with the intrinsic 
delay introduced by `twisting' a photon.
The authors reach the surprising result that the addition of OAM reduces the 
delay (i.e., makes the pulse somewhat faster) with respect to exactly the same 
beam with no OAM.
Finally, a recent paper \cite{Sirtsuga} analyzes theoretically how to 
set the group velocity of ultrashort light pulses in vacuum to arbitrary values 
within the focal region.

As distinct from focused beams, pulses of (pseudo)non-diffracting beams are
propagation-invariant, i.e., their intensity profile changes 
neither in any lateral nor in the axial direction over a large spatial range.
Experimental realizability of the simplest type of such pulses, called the
Bessel-X pulse, was demonstrated in \cite{PRLmeie}, where we used the
same optical scheme with an annular slit as Durnin \textit{et al.} in their
seminal work \cite{Durnin} on the Bessel beam.
A narrow annular slit with angular radius $\theta $ ensures that the 
frequency-dependent phases of all Bessel beam constituents of the
polychromatic field are proportional to the frequency:
$z\cos \theta \thinspace $$\omega /c$.
The group velocity in this case is $c/\cos \theta $, i.e., superluminal, since 
it is given by the reciprocal of the mixed derivative of the phase with respect to 
frequency and propagation coordinate $z$.
There is  massive literature on various non-diffracting pulsed waves (also 
called propagation-invariant localized waves) which are classified into 
superluminal, luminal, and subluminal types (see \cite{LW2} and reviews 
\cite{DonelliSirged,revPIER,revSalo,MeieLorTr,KiselevYlevde}).

In a sense, the discussion of \textit{subluminality} of structured light in the 
recent literature is \textit{d{\'e}j{\`a} vu}: in the preceding decade the 
meaning of  \textit{superluminality}  of non-diffracting pulses was intensively 
debated.
Despite the experimental proofs \cite{PRLmeie,exp2,exp3}, the 
feasibility of superluminal group velocities of Bessel-X-type pulses 
was questioned until we carried out direct measurements
\cite{meieXfemto,meieOPNis} of the spatio-temporal electric field of such pulses
generated by a refractive axicon (conical lens).

In contradistinction to superluminal non-diffracting pulses generated by an 
annular-slit-and-lens scheme or a refractive or reflective axicon, the 
so-called pulsed Bessel beams generated by circular diffraction gratings---which 
are precisely what were considered in \cite{Gio} and \cite{Alf}---are always 
subluminal and are not propagation-invariant because they spread temporally.
This was shown theoretically a long time ago 
\cite{LiuPulsedBB,DifRefAxiconBB,PorrGaussjaPBB} and experimentally in 
\cite{meieDifAxicon0,meieDifAxicon1} with femtosecond-range temporal and 
micrometer-range spatial resolution of the propagating field.
A mistake in mathematical derivation of group velocities, which led to 
inability of some authors to distinguish the Bessel-X pulse from the pulsed 
Bessel beam and to accept the superluminality of the former, is analyzed 
in the review article \cite{Minupeatykk}.

In some forthcoming calculations we resort to two-dimensional analogues of 
cylindrically symmetric 3D fields, i.e., to fields that depend---in addition to the longitudinal coordinate and time---on only one transverse coordinate. 
In expressions for such 2D fields the zeroth-order Bessel 
function $J_{0}$ is replaced by a cosine. Propagation properties of 
2D field pulses coincide, \textit{mutatis mutandis}, 
with those of 3D ones.
For a propagation-invariant non-diffracting pulse, the so-called focus wave 
mode which was theoretically intensively studied in the end of the previous 
century, this coincidence was shown in \cite{FWMmeieExp}, where experimental 
feasibility of this exactly luminal wideband pulse had been also demonstrated 
for the first time.
A general theory of 2D luminal and superluminal propagation-invariant 
(localized) waves has been developed in \cite{Yannis2D}.
Having in mind, e.g., the fast development of light-sheet microscopy and emerging applications of Airy beams, the 2D non-diffracting pulses cannot be considered as inferior to the 3D ones.
A very recent article \cite{Xsheet} demonstrates the generation 
of 2D non-diffracting pulses of different group velocities by means of a 
spatial light modulator.

For the forthcoming analysis it is useful to recall that even the pulses of the 
fundamental Gaussian beam may exhibit both slightly superluminal and slightly 
subluminal propagation near their focus or  the Rayleigh range 
\cite{HoGouy0,HoGouy,isodiffrWinful,Minu3Gaussi,PorrGaussNega}.
The behavior of the group velocity of a polychromatic beam depends on the frequency 
dependence of the (interrelated) parameters of the monochromatic constituent 
beams, which in turn is determined by the optics generating the beam.
On the basis of the character of the frequency dependence, pulses built from 
Gaussian, Bessel, Airy, etc  beams can be generally divided into different 
types \cite{Minu3Gaussi,3tyypi,minuAiry,meieAiry4tyyp}, each possessing 
its own specific group velocity properties.

If we juxtapose the recent papers \cite{Gio,Horv2015,Bareza,Bouch,MinuComm1,MajoranaTorn,Alf,MinuComm2,Faccio2}
with the earlier literature referred to above, the following questions arise.
How is the group velocity that is evaluated in an ordinary way related to the propagation 
time (time of flight) of the pulse between the input (source) and output 
(recording) planes?
How does one take into account non-constancy of the velocity over 
the propagation distance?
Is the axial projection of the velocity or some other 
quantity appropriately related to the times or delays in flight, measured in 
experiments with hollow beams---such as the Laguerre-Gauss and other beams with 
OAM?
And last but not least, how does one resolve the contradiction between the 
notion of the spatially averaged group velocity---which is always subluminal and 
was introduced in \cite{Gio} for relating to time-of-flight experiments---and the superluminality of X-type non-diffracting pulses?
The purpose of the present study is to give answers to these questions based on a model pulsed beam---the Bessel-Gauss pulse.
This model pulse seems to be the most appropriate one for our purpose since it exhibits both superluminal and hollow-beam propagation stages and is physically realizable.

The paper is organized as follows.
In the next section, different definitions of group velocities at off-axis field 
points are considered and the most suitable quantity for relating to the axial time 
of flight is chosen using a cylindrically symmetric Bessel-Gauss pulse 
\cite{BessGauss0,BessGaussPulse,PorrBessGauss,BessGaussMeie} as a numerical test.
Having in mind that the notion of group velocity generally describes well the 
propagation of narrow-band pulses only, in Sec. III spatio-temporal evolution of
an ultrashort Bessel-Gauss pulse with a particular wide-band spectrum is
calculated resulting in a series of 3D plots.
These plots can be considered as ``snapshots in flight'' of the pulse and they 
are analyzed in order to verify and to more deeply interpret the results of 
Sect. 2.
Finally, Sec. IV is devoted to solving the paradox that the formula of spatially
averaged group velocity derived in \cite{Gio}, which supposedly is equal to the
ratio of propagation distance and time, does not apply to superluminal pulses.

\section{Differently determined group velocities for example of Bessel-Gauss 
pulse}

The well-known expression of the group velocity at a point $\mathbf{R}$ of a 
three-dimensional wavepacket with the carrier (mean) frequency $\omega $ reads 
\cite{BornWolf}
\begin{equation}v\left (\mathbf{R}\right ) =\frac{1}{\left \vert 
grad\left [\varphi _{\omega }^{ \prime }\left (\mathbf{R}\right )\right ]\right 
\vert } , \label{BW}
\end{equation}
where $\varphi _{\omega }^{ \prime }\left (\mathbf{R}\right )$ 
denotes the derivative of the spatial phase of the monochromatic constituents 
of the packet with respect to frequency.
From an experimentalist's point of view the observable quantity of primary 
interest is time $\tau $  that the pulse peak (or other feature) needs to travel 
from a certain starting (source) plane to the output (recording) plane.
In the case of a plane-wave pulse or when the pulse peak propagates along a 
straight line---the optical axis $z$ of a paraxial beam---it is not a problem 
to relate $v(\mathbf{R})\text{,
}$ the time $\tau $, and the propagation depth $z =z_{out} -z_{in}$ with each 
other.
But generally, even for paraxial beams, the quantity $v(\mathbf{R})$ is 
inappropriate for finding the travel time $\tau $ for a given propagation depth.
For example, the intensity of a Laguerre-Gaussian beam is concentrated not on the 
axis but at a certain radial distance from it which increases as the pulse moves 
away from the focus.
Due to the latter circumstance, as pointed out in \cite{MinuComm1}, the pulse 
travel time $\tau $ is not determined by $v\left (\mathbf{R}\right )$ which is 
the magnitude of the  group velocity vector, but rather by its projection onto 
the propagation axis $z$.
Also, it should be stressed here that this $z$-projection is generally not 
given by replacing the gradient operator in Eq.~(\ref{BW}) with $ \partial / 
\partial z$.

In order to propose such velocity quantities which are appropriate for relating 
to travel time versus propagation depth data from time-of-flight-type 
experiments, let us take one step back in the derivation of Eq.~(\ref{BW}).
As shown in \cite{BornWolf}, the instant of time at which the modulus of the 
field reaches its local maximum at position $\mathbf{R}$ is given 
by
\begin{equation}\tau (\mathbf{R}) =\varphi _{\omega }^{ \prime }\left 
(\mathbf{R}\right ) .
\label{Delay}
\end{equation}
If the pulse peak, i.e., its absolute maximum passes through the point 
$\mathbf{R}$, the quantity $\varphi _{\omega }^{ \prime }\left (\mathbf{R}\right 
)$  directly measures the time of flight of the pulse.
For definiteness, let us choose the origin of the coordinate $z$ so that the 
pulse peak passes through the plane $z =0$  at the instant $t =0$.
Then, we can define an average group velocity in the direction of the optical 
axis over the distance $z$ as follows:
\begin{equation}v_{a}(z ,r ) 
=\frac{z}{\tau  (z ,r )} .
\label{Va}
\end{equation}
Here, we have introduced cylindrical coordinates $(z ,r ,\phi )$ 
for the point $\mathbf{R}$ and assumed that $\varphi _{\omega }^{ \prime }\left 
(\mathbf{R}\right )$ does not depend on the azimuthal angle $\phi $, which is 
true not only for beams with cylindrical symmetry but also for beams possessing 
orbital angular momentum, as long as the contribution of the azimuthal angle to 
the phase of their field does not depend on the frequency.
In the supplemental material of Ref.
\cite{Gio} this velocity has been obtained by harmonic averaging of $v\left 
(\mathbf{R}\right )$ over distance from $z_{1}$ to $z_{2}$ and used as an 
intermediate quantity in the derivation of a three-dimensionally averaged velocity 
which we consider in Sec. IV.

The Born-Wolf velocity given by Eq.~(\ref{BW}) in the cylindrical coordinates 
reads\begin{equation}v(z ,r) =\frac{1}{\sqrt{\left [\frac{ 
\partial }{ \partial z}\tau  (z ,r )\right ]^{2} +\left [\frac{ \partial 
}{ \partial r}\tau  (z ,r)\right ]^{2}}} .
\label{V}
\end{equation}
As this quantity is the magnitude of the group velocity vector which has been 
directed along $grad\left [\tau (z ,r)\right ]$, the projection of the group 
velocity onto the optical axis is obtained through multiplication by the 
directional cosine resulting in
\begin{equation}v_{z} (z ,r ) =\frac{\frac{ \partial }{ \partial 
z}\tau (z ,r)}{\left [\frac{ \partial }{ \partial z}\tau  (z ,r
)\right ]^{2} +\left [\frac{ \partial }{ \partial r}\tau  (z ,r)\right 
]^{2}} \label{Vz}
\end{equation}and, in the same way, the radial component of the velocity 
reads\begin{equation}v_{r} (z ,r) =\frac{\frac{ \partial }{ 
\partial r}\tau (z ,r)}{\left [\frac{ \partial }{ \partial z}\tau  (z 
,r )\right ]^{2} +\left [\frac{ \partial }{ \partial r}\tau  (z ,r
)\right ]^{2}} .
\label{Vr}
\end{equation}In the case $\frac{ \partial }{ \partial r}\tau (z ,r ) 
=0$, e.g., for on-axis points, the harmonic averaging of the axial group 
velocity $v_{z} (z ,r )$ results in $v_{a}(z ,r )$ 
defined by Eq.~(\ref{Va}).

\begin{figure}
\centering
\includegraphics[width=8.5cm]{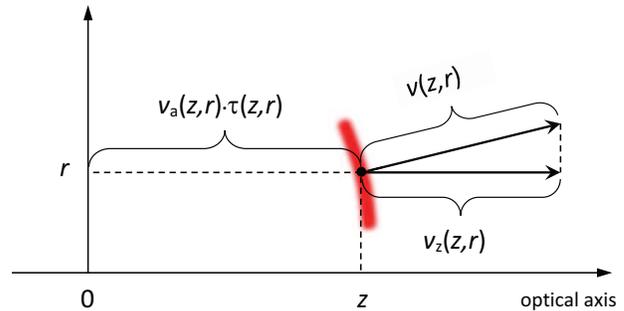}%
\caption{\label{FigVskeem}
Scheme of distinctions between the group velocities defined in the text.}
\end{figure}

Figure 1 \ref{FigVskeem} illustrates the differences between the quantities defined by 
Eqs.~(\ref{Va})-(\ref{Vz}) in the case of an off-axis (hollow) diverging pulse.

Next, we calculate these velocities for a particular test field.
As the latter we have chosen a pulse of the Bessel-Gauss beam, since it is a 
paraxial, Gaussian-apertured, finite-energy and, hence, physically realizable  
version of the Bessel-X pulse.
Moreover, it exhibits a pronounced and independently adjustable superluminality 
in the focal region and in the far field its intensity is concentrated in a 
ring of diverging radius---quite similarly to pulsed Laguerre-Gauss beams.
The Bessel-Gauss beams---the monochromatic constituents of the pulse---are 
tractable as superpositions of Gaussian beams (modes), the optical axis of each 
of which lies along a generatrix of a cone.
Half of the cone apex angle---which is called the axicon angle and typically 
designated by $\theta$---determines the superluminal group velocity of the 
Bessel-X pulse as equal to $c/\cos \theta$ and, consequently, the same 
velocity for a pulse in the focal region of the Bessel-Gauss beam 
(henceforth---the BG \ pulse).

It is known that the behavior of the group velocity in a pulsed Gaussian beam is 
determined by the frequency dependence of the (interrelated) parameters of the beam 
\cite{HoGouy0,HoGouy,isodiffrWinful,Minu3Gaussi}.
For example, if the Rayleigh range is inversely proportional to frequency, the 
on-axis group velocity is slightly superluminal in the focal region.
The same holds for the Laguerre-Gauss beams \cite{Bouch}.
Conversely, if the dependence is proportional to the frequency, the group velocity is 
slightly subluminal there.
In order not to mix these effects on the group velocity with 
axicon-angle-controlled superluminality, we chose Gaussian beams whose Rayleigh 
range $z_{R}$ is frequency independent to play the role of constituents of the 
BG pulse.
Pulses formed from such beams---isodiffracting pulses---possess strictly 
luminal (equal to $c$) group velocity along the whole propagation axis 
\cite{isodiffrWinful,Minu3Gaussi}.
Also, vanishing derivative $ \partial z_{R}/ \partial \omega $  keeps group 
velocity expressions comparatively simple in the case of isodiffracting pulses.

\begin{figure}
\centering
\includegraphics[width=8.7cm]{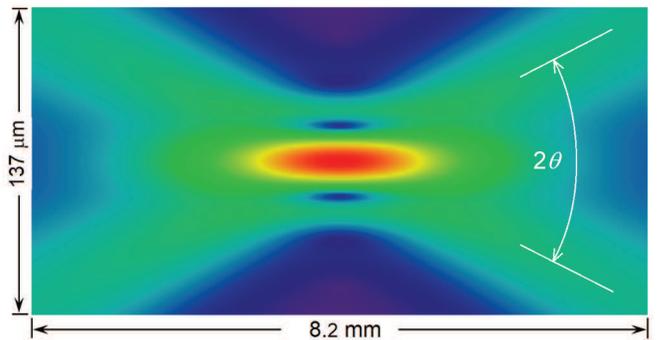}%
\caption{\label{FigSqrtMoodulColor}
(Color online) Illustrative plot of a Bessel-Gauss beam.
Depicted is the square root of the modulus of Eq.~(\ref{BGE}) without the first 
fraction.
Note that the scale of the vertical axis representing the transverse coordinate 
$\pm r$ has been magnified 60 times relative to that of the $z$ axis.
Beam parameters: Rayleigh range $z_{R} =3mm$, wavelength $\lambda  =2\pi /k 
=0.718$ $\mu m ,$ axicon angle $\theta  =1^{\circ} ,$ divergence of the constituent 
Gaussian beams  $\theta _{0} =\left (kz_{R}/2\right )^{ -1/2} =0.5^{\circ}$.}
\end{figure}

According to \cite{PorrBessGauss} the monochromatic Bessel-Gauss wave function 
(without the time-dependent factor $\exp i\omega t$) of wavenumber $k =\omega 
c$ reads
\begin{multline}\psi (z ,r ,k) =\frac{iz_{R}}{z +iz_{R}}
J_{0}\left (\frac{iz_{R}}{z +iz_{R}}\theta kr\right ) \\
 \times \exp 
\left \{ -ik\left [\frac{r^{2} +z^{2}\theta ^{2}}{2\left (z +iz_{R}\right )} +z\left (1
 -\frac{\theta ^{2}}{2}\right )\right ]\right \} , \label{BGE}\end{multline}
where $J_{0}$ is the zeroth-order Bessel 
function of the first kind and $\theta $ is the Axicon angle (apex half-angle of 
a cone over the surface of which the directions of constituent Gaussian beams 
are evenly distributed).
Equation~(\ref{BGE}) and Fig.~\ref{FigSqrtMoodulColor} exhibit the wave function 
dependencies characteristic to both the Gaussian and Bessel beams.

\begin{figure}
\centering
\includegraphics[width=8.7cm]{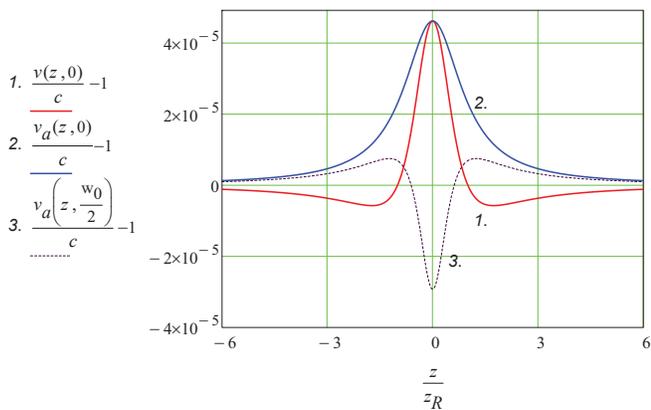}%
\caption{\label{FigTelgjoonel3kiirust} (Color online) 
Curves 1 and 2: on-axis group velocities defined by Eq.~(\ref{V}) and 
(\ref{Va}), respectively, shown as normalized velocity changes vs the 
normalized propagation distance $z$.
Curve 3: averaged group velocity along a line shifted radially from the  
$z$ axis to half of the beam waist radius $w_{0} =0.026$ mm or---in 3D terms---on 
the surface of a cylinder with radius $w_{0}/2 =0.013$ mm.
Beam parameters are the same as in Fig. \ref{FigSqrtMoodulColor}, but the axicon angle $\theta  =0.55^{\circ}$ is smaller.}
\end{figure}

The Gaussian radial profile suppresses the Bessel-function radial profile in 
the vicinity of the focal plane $z =0$ and we see the side-maxima of the Bessel 
function thanks to choosing the axicon angle twice larger than the divergence 
of the constituent Gaussian beams and taking the square root of the modulus (for 
improving contrast of the image).
The spatial dependence of the field outside the focal region, i.e., where $\vert 
z\vert  >z_{R}$ is seen with noticeable intensity thanks to omitting the factor 
$iz_{R}/($$z +iz_{R})$ in Eq.~(\ref{BGE}).
This factor does not contribute to $\tau (z ,r)$ according to Eq.~(\ref{Delay}) 
since $z_{R}$ is frequency-independent.
The exponential factor contributes to $\varphi _{\omega }^{\prime }\left 
(\mathbf{r}\right)$ by the real part of the square brackets (divided by $c ,$ 
since $ \partial / \partial \omega  =c^{ -1} \partial / \partial k$).
The derivative of the phase of the Bessel function can be evaluated 
\textit{via} the identity
\begin{equation} \frac{ \partial }{ \partial \omega 
}\arg U =\frac{1}{c}\ensuremath{\operatorname*{Im}}\left (\frac{1}{U}\frac{ 
\partial U}{ \partial k}\right ).
\end{equation} Expressing the derivative of the zeroth-order Bessel function 
through the first-order one and carrying out some algebra, we obtain
\begin{multline}\tau (z ,r) =\frac{z}{c} +\frac{1}{c}\frac{z\left (r^{2} 
-\theta ^{2}z_{R}^{2}\right )}{2(z^{2} +z_{R}^{2})} + \\
 +\frac{1}{c}\ensuremath{\operatorname*{Im}}\genfrac{\{}{\}}{}{}{i\theta 
rz_{R}\thinspace J_{1}\genfrac{(}{)}{}{}{i\theta rz_{R}k}{z +iz_{R}}}{\left (z 
+iz_{R}\right )\thinspace J_{0}\genfrac{(}{)}{}{}{i\theta rz_{R}k}{z +iz_{R}}} 
, \label{BGDelay}\end{multline}where $k$ now is the mean wavenumber of a 
quasimonochromatic wave packet.
The first term on the right-hand side gives the arrival time at point $z$ of a 
$\delta $-shaped plane wave pulse started from the origin
$z =0$ at $t =0.$ Hence, the following two terms are responsible for the 
temporal shift of the BG pulse relative to the ``signal'' pulse.
It follows from Eq.~(\ref{BGDelay}) that everywhere on the axis $z$ the shift 
is negative, since with $r =0$ the third term vanishes.
Negative temporal shift means superluminal group velocity.
Indeed, as seen in Fig. \ref{FigTelgjoonel3kiirust}, the average group 
velocity $v_{a}$ ranges from the superluminal value $c/(1 
-\theta ^{2}/2)$ $ \simeq c/\cos \theta $  (which is the constant superluminal 
group velocity of the Bessel-X pulses) at the origin down to $c$ at large 
propagation distances.
This result was obtained earlier in \cite{PorrBessGauss}, although without 
pointing out that this is the averaged group velocity, not the common local 
(Born-Wolf) group velocity $v$.
In Fig. \ref{FigTelgjoonel3kiirust} the latter exhibits transition to
subluminal values as soon as 
the point exits from the Ra eigh range and reaches its minimum at $z = 
\pm \sqrt{3\ }z_{R}$ irrespective of the axicon angle $\theta $.
It is remarkable that the on-axis group velocity of such ordinary Gaussian 
pulsed beam, whose divergence $\theta _{0}$ is frequency-independent but the 
Rayleigh range is reciprocally proportional to frequency, behaves exactly the 
same way.
A mathematical reason for this amazing coincidence is the identity of the 
expressions for $\tau (z ,0)$ in both cases if one replaces the angle $\theta $ 
with $\theta _{0}$.\  Physically, while the superluminality of the juxtaposed 
Gaussian beam pulse is caused by the frequency-dependent Gouy phase $\arctan 
(z/z_{R})$, the superluminality of the BG pulse is a result of interference 
between its luminal Gaussian beam pulse constituents propagating under the 
axicon angle.

The third curve in Fig.~\ref{FigTelgjoonel3kiirust} shows that while the averaged
velocity is superluminal 
everywhere on the propagation axis, it turns out to be subluminal at off-axis 
points in the focal region.
The Born-Wolf group velocity $v$ along the same off-axis line is subluminal 
everywhere and exhibits two subluminal minima in the focal region---the curve 
being quite similar to that for the Laquerre-Gauss pulsed beam (Fig.
\ref{FigVskeem} in \cite{MinuComm1}) and to curve \ref{FigVskeem} in Fig.
5 below.

\begin{figure}
\centering
\includegraphics[width=8.7cm]{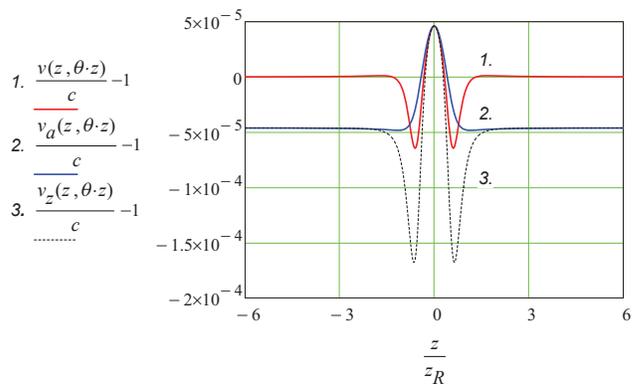}%
\caption{\label{FigKaldjoonel3kiirust} (Color online) 
Group velocities defined by Eq.~(\ref{V}), (\ref{Va}), and (\ref{Vz}) 
on the conical surface given by $r =\theta z$, plotted  as functions of the 
propagation distance (curves 1, 2, and 3, respectively).
Beam parameters are the same as in Fig. \ref{FigTelgjoonel3kiirust}.
}
\end{figure}

Unless $\theta  \ll \theta _{0}$, in other words, unless the BG pulse is almost 
like a Gaussian beam pulse, the on-axis and near-axis intensity is very low 
outside the Rayleigh range and therefore the curves in
Fig.~\ref{FigTelgjoonel3kiirust} are not of our main interest.
According to our goals formulated in Sec. I, we must study the group velocity 
behavior on the peak of the pulse.
For the BG pulse, it means asymptotically along a generatrix of the cone, 
i.e., along the straight line $r =\theta \thinspace z$, see Fig.
\ref{FigKaldjoonel3kiirust} (being precise, the peak of the pulse transposes
itself from the optical axis 
to the surface of the cone $r =\theta \thinspace z$ when $z >z_{R}$ as we will see 
in the next section).

We see that the Born-Wolf group velocity is luminal everywhere in the  Rayleigh 
range.
This is understandable, because at large distances the pulse peak area constitutes 
a converging (when $z < -z_{R}$) or diverging (when $z >z_{R}$) spherical zone 
and it is well known that the velocity of a spherical wave equals $c$.
At the same time, since the group velocity vector is directed under the axicon 
angle with respect to the $z$ axis, its projection $v_{z}$ to the propagation 
axis is \textit{subluminal} everywhere in the  Rayleigh range (as pointed out 
also in \cite{MinuComm1,MajoranaTorn}).
The same holds for the average group velocity $v_{a}$ which is directly related 
to the delays that are measurable experimentally.
Asymptotically $v_{z} \approx v_{a} \approx c\cos \theta $.
Hence, in the case of pulses whose maximum does not propagate along the optical 
axis (Laguerre-Gauss pulsed beams and alike, the BG pulse outside the Rayleigh 
range), if one studies the behavior of their group velocity \textit{via} 
measuring their arrival delays in the far field, the results are determined by 
the subluminal plateau.
It means that in a common experimental geometry this plateau extends over 
distances much larger than the focal region where the velocity variations take 
place and, as a result, the variations remain masked in the delay data.
In order to study the interesting behavior of the group velocity in the focal 
region, which distinctively depends on the type and parameters of the beam, in 
addition to the temporal resolution one must apply at least a sub-millimeter 
spatial resolution in the vicinity of the focus.
This can be accomplished by the spatially encoded arrangement for temporal analysis by dispersing a pair of light E-fields (SEA TADPOLE) technique, as was done in 
Ref.~\cite{meieXfemto}.
These conclusions constitute one of the main results of the present study.

We point out that for plots in Figs.~3-5
the angles $\theta $  and $\theta _{0}$ were taken almost equal.
Changing the ratio $\theta /\theta _{0}$ reveals the following: (i) the region 
of the variable behavior of the velocities is confined by $ \pm z_{0}$ rather 
than by $ \pm z_{R}$, with $z_{0}$ being the $z$ coordinate of the point where the 
line $r =\theta \thinspace z$ attains the radial distance of the first zero of 
the Bessel profile; (ii) if $\theta /\theta _{0} >2$ the minima of $v$ and 
especially of $v_{z}$ become very deep in the vicinity of $ \pm z_{0}$ while 
the radial component $v_{r}$ acquires large values ( $v_{r} >0.5c$ when $\theta 
/\theta _{0} >3$) at these locations.
Such steep variations start to manifest themselves also at the locations where 
the line reaches the radial distances of the next zeros of the Bessel profile.
Although the beam intensity vanishes completely only on the rings corresponding 
to the zeros of the Bessel profile  in the plane $z =0$, at larger distances 
these zeros show up as intensity minima if $\theta /\theta _{0} >2$.
Thus, the intensity minima cause the steep increase of the radial component 
$v_{r}$ of the group velocity.

\begin{figure}
\centering
\includegraphics[width=8.7cm]{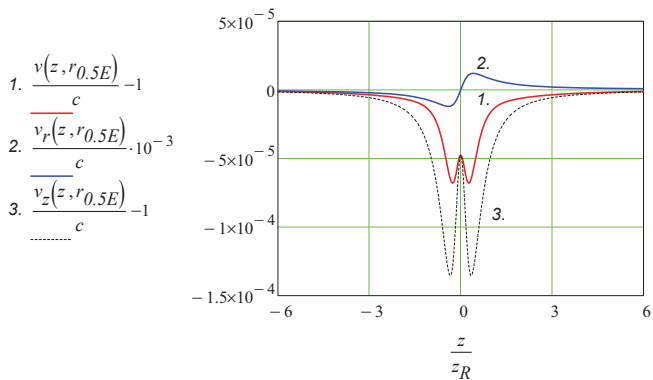}%
\caption{\label{FigPooleEraadiusel3kiirust} (Color online) 
Magnitude of the group velocity vector and its radial and axial components 
(curves 1, 2, and 3, respectively) on the surface of a cylinder with radius 
$r_{0.5E} =0.014\ $mm $\approx w_{0}/2$.
Beam parameters are the same as in Fig. \ref{FigTelgjoonel3kiirust}.
Note the scaling of $v_{r}$.
}
\end{figure}

In order to comprehend the behavior of the radial component $v_{r}\left (z 
,r\right )$, we studied its $z$-dependence at fixed non-zero values of $r$  and 
its $r$-dependence at fixed non-zero values of $z$.
Of course,  $v_{r}(z ,0)$ and
$v_{r}(0 ,r)$ are identically equal to zero as follows from symmetry 
considerations already.
Figure \ref{FigPooleEraadiusel3kiirust} shows the $z$ dependence of the magnitude of the group velocity and its 
components on a cylindrical surface whose radius is equal to the HWHM of the pulse 
modulus at origin.

It follows from Fig. \ref{FigPooleEraadiusel3kiirust} that the plotted
off-axis velocities $v$  and $v_{z}$ are subluminal 
everywhere and have obtained two minima (as is the case with the off-axis group 
velocity of Laguerre-Gauss beams \cite{MinuComm1}).
Since the equality $v_{z}^{2} +v_{r}^{2} =v^{2}$ holds for any point, the 
minima coincide with the maxima of $\vert v_{r}\vert $.
When $z <0$, the beam converges and, therefore, $v_{r} <0$ at that stage of the 
propagation of the pulse.
At the diverging stage the group-velocity vector has been directed away from 
the propagation axis, i.e., $v_{r} >0$ when $z >0$.
However, there is a subtlety: at a fixed distance $0 <z \ll z_{R}$, the 
positive value of $v_{r}$ grows steeply (up to $0.5c$) when $r$ approaches the 
radial distance $r_{01}$ of the first zero of the Bessel profile, then at the 
point $r_{01}$ reverses the sign of its value and decreases to zero at the 
half-way to the radial distance $r$\textsubscript {$02$} of the next zero, where 
the the same behavior repeats itself.
Hence, figuratively speaking, the cylindrical surfaces with radii $r_{01} 
,r_{02} ,\ldots $ attract the field flow and deflect the group velocity vector 
\ in the vicinity of the focus.

In the next section, we correlate the obtained velocity curves with numerically 
simulated propagation of a wideband BG pulse whose duration and other 
parameters are suitable from an experimentalist's point of view.
At the same time we recall that the notion of group velocity assumes neglecting 
the higher than $\varphi _{\omega }^{ \prime }\left (\mathbf{R}\right )$ 
derivatives of the spatial phase, which become significant with increasing 
bandwidth and are responsible for distortions of the pulse in the course of 
propagation.

\section{Propagation of model Bessel-Gauss pulse with Poisson-like spectrum}

For a better understanding of the results of the preceding section, it would be 
helpful to graphically depict the spatiotemporal evolution of a BG pulse.
Porras \cite{PorrBessGauss} has found a suitable closed-form expression for
the wave function of a BG pulse with the so-called Poisson-like spectrum $f 
(\omega ) =$ $\pi  t_{0}^{n +1} \omega ^{n} \exp  ( -\omega  t_{0})/n !$ , 
where $\omega  >0$, $t_{0} >0$ determines the duration of the pulse and $n$ is 
a natural number.
With somewhat changed designations the expression reads 
\begin{multline}\Psi (z ,r ,t) =\frac{i z_{R}}{z +i z_{R}} 
\left \{\frac{i t_{0}}{\sqrt{(t_{c} +i t_{0})^{2} -[\frac{i z_{R}}{c 
(z +i z_{R})} \theta  r]^{2}}}\right \}^{n +1} \\ \times P_{n} 
\left \{\frac{t_{c} +i t_{0}}{\sqrt{(t_{c} +i t_{0})^{2} -[
\frac{i z_{R}}{c (z +i z_{R})} \theta  r]^{2}}}\right \}\text{,} 
\label{Pulse}\end{multline}
where $P_{n} ()$ is the Legendre polynomial of the 
order $n$ and 
\begin{equation}t_{c} =t -z (1 -\theta ^{2}/2)/c -\frac{1}{2 c (z +i z_{R})} 
(r^{2} +z^{2} \theta ^{2})
\end{equation}is a space-dependent complex time.
As is well known, the group delay and group velocity expressions are, 
strictly speaking,
meaningful if the FWHM $ \Delta \omega $ of the pulses' spectrum is much smaller than 
its mean frequency $\omega _{m}$.
If this narrow band condition is not fulfilled, the pulse undergoes distortions 
in the course of propagation.
On the other hand, a
narrow-band BG pulse would inevitably be too long for observation of the 
expected tiny differences of
its propagation velocity from that of a plane wave, i.e., from $c$.
Besides, since it follows from Eq.\nolinebreak\relax (\ref{Pulse}) that  $ 
\Delta \omega /\omega _{m} \sim 1/\sqrt{n}$ , the narrow band condition would 
require a numerical evaluation of Legendre polynomials of very high order, 
which could cause computational
problems.
We have chosen $n =16$ as this value corresponds roughly to $ \Delta \omega 
/\omega _{m} \lesssim 1/2$ and gives a three-cycle pulse.
Last but not least, temporal shifts as small as $ \approx 1$ fs of 
such ultrashort light pulses are easily observable with our interferometric 
set-up based on a supercontinuum laser \cite{meieDifAxicon1,meieAiry4tyyp}.

\begin{figure*}
\centering
\includegraphics[width=17.5cm]{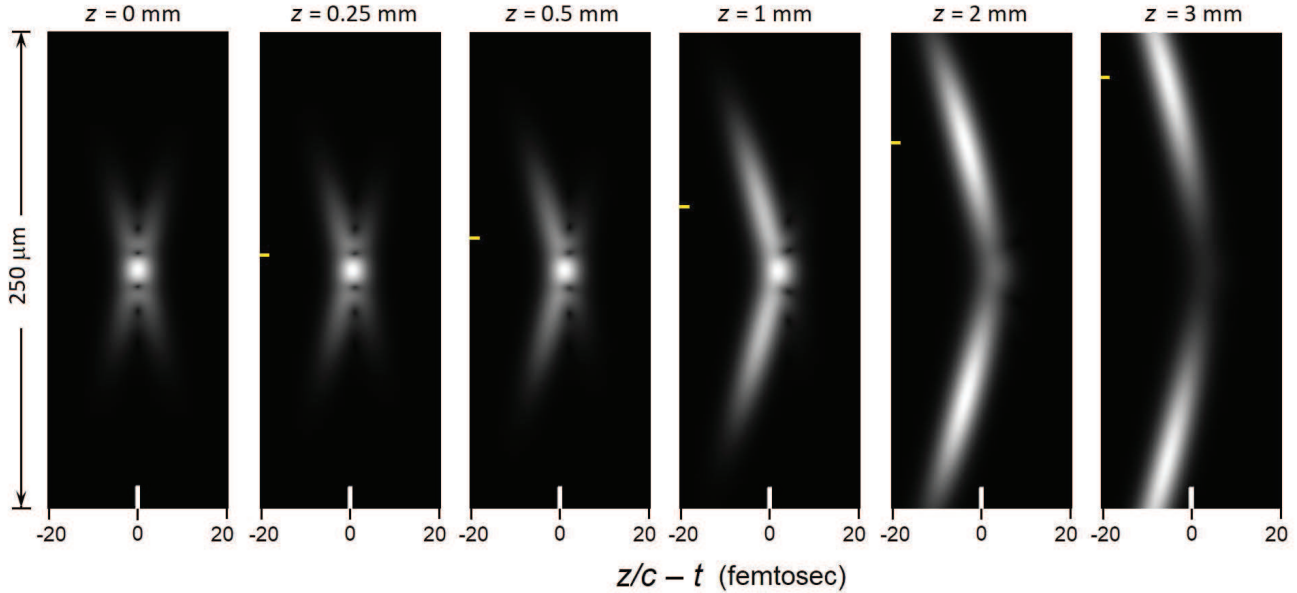}%
\caption{\label{FigBXevolution}
Evolution of the modulus of the BG pulse plotted from Eq.~(\ref{Pulse}) at 
sequential distances from the focus at $z =0$.
The vertical axis represents the transversal coordinate $ \pm r$.
 The vertical white bars indicate the positions of $t =z/c$, i.e., the time 
instants that an ultrashort plane-wave pulse would reach the corresponding 
coordinate $z$ if it co-propagated along the $z$ axis together with the BG 
pulse.
The yellow horizontal dashes mark the radial distances $r =\theta z$.
The parameters of the pulse: the axicon angle $\theta  =2^{\circ}$, the Rayleigh 
range $z_{R} =3$ mm, $t_{0} =$ 7 fs ($ct_{0} =2$ $\mu$m), and $n =16$.
The frames have been normalized to unity (white).
For a better revealing of the weak wings of the pulse, the grayscale has been 
taken proportional to $\sqrt{\vert \Psi (z ,r ,t)\vert \text{.}}$
}
\end{figure*}

Evolution of the pulse in the course of propagation is shown in 
Fig.~\ref{FigBXevolution}.
The plots depict radial and temporal behavior of the pulse modulus in six 
cross-sectional planes with fixed values of $z$.
A part of the profile seen to the left from the abscissa value $z/c -t =0$ 
shows the temporal  behavior of the modulus in every given $z$ plane after the 
instant when a luminally propagating signal would cross the $z$-plane (it is 
assumed that the ``signal'' starts when the peak of the pulse is at the position 
$z =0$).
Conversely, a part of the profile seen at the positive values of $z/c -t$ shows 
the temporal  behavior of the modulus ahead of the signal.
Such a representation of spatio-temporal evolution of ultrashort light pulses 
is common in femtosecond-resolution measurements that use the SEA\ TADPOLE 
technique.
However, it is easier to comprehend the plots as  ``snapshots in flight'' or 
``still frames'' taken in the meridional plane of the beam at sequential fixed 
time instants.
In the given case the instants would be  $t =0 ,$ $t =834$ fs, $t =1.6$ ps, $t 
=3.3$ ps, $t =6.7$ ps, and $t =10$ ps (the values corresponding to  $z =0 ,\ 
0.25 , .
..3$ mm) and the horizontal extent of each ``frame'' is $12.5$ $\mu m$.
Such an equivalence of the two representations is possible because the paraxial 
and ultrashort field practically does not  change its shape when propagating 
over a distance as short as $12.5\ \mu m$ during $ \approx 40$ fs.
Indeed, to check the equivalence we also calculated the plots from Eq.~(\ref{Pulse}) 
with the fixed values of the variable $t$ and it turned out that the 
values of the modulus in corresponding matrices (every element or pixel) of the 
two sets of plots are equal with accuracy $ \leq 0.5 \%$.
In what follows,  we refer to Fig.~\ref{FigBXevolution} as if it contains the ``still frames.''

The first frame depicts the instantaneous intensity profile (more exactly, its 
fourth root for better visibility of low-intensity features) of the BG pulse at the
origin (in the focus of its Gaussian constituents).
The X-like (actually a double cone like in 3D space) structure peculiar to the 
Bessel-X pulse \cite{PRLmeie, meieXfemto} with some residual Bessel beam 
minima around the apex is clearly seen.
We have intentionally chosen the axicon angle $\theta  =2^{\circ}$ larger than that 
in the plots of the previous section, i.e., four times exceeding the Gaussian 
divergence $\theta _{0} =0.5^{\circ}$, in order to reveal these minima and the 
wings of the pulse.
The next two frames indicate that the pulse is superluminal---it outstrips a 
copropagating plane wave reference whose sequential positions have been marked 
by the small white vertical bars.
As distinct from the propagation invariance of the Bessel-X pulse, the BG pulse 
in the subsequent frames loses its double-conical shape with superluminal apex 
due to the limited (Gaussian) transversal dimension of the constituent beams.
As a result, after reaching the Rayleigh range the BG\ pulse peak intensities 
constitute a ring whose radius grows linearly with propagation distance.
In other words, points of highest intensity move along the cone surface, the 
generatrix of which obeys the equation $r =\theta z$ used in plots of the 
preceding Section.
The last three frames show that the $z$-directed velocity of propagation of the 
ring is subluminal thus indicating the reason for the subluminal plateau in Fig. \ref{FigKaldjoonel3kiirust}.

Horizontal dashes in Fig. \ref{FigBXevolution} mark values of $r$ at 
which the line $r =\theta z$ intersects the frames.
Note that the abscissa scale is stretched 20 times with respect to the $r$ axis 
of the frames.
This scale stretching explains why, in the last three frames, the pulse fronts 
are not practically vertical, i.e., perpendicular to the line $r =\theta z$ (as 
they should be in the case of equal scales).

The main conclusions from the plots of the pulse propagation are as follows.
\begin{enumerate}
\item Group velocities are superluminal in the spatial region where the 
constituent Gaussian beam pulses, propagating along a pair of the cone 
generatrices of opposite inclination with respect to the $z$  axis,  overlap and 
interfere with each other.
Note that the pulse profile in the first two frames resembles the Bessel-X 
pulse which is superluminal (the similarity of the Bessel-Gauss and Bessel-X 
fields has also been demonstrated in \cite{PorrBessGauss}).
\item Outside this region---see the last two frames---the pulse propagates 
like a spherical zone and although the group-velocity vector in the peak of the 
pulse is directed along the generatrix and its magnitude is $c$, the projection 
of the peak position onto the $z$  axis lags behind the luminal signal.
Figure \ref{FigKaldjoonel3kiirust} indicates that this lag at large propagation
depths $z \geq z_{R}$ 
results from subluminal values of $v_{a}$ and $v_{z}$ at the radius of the peak 
position, which both approach their far-field value $c\cos \theta $.
\item The two middle frames show that in the transition region the superluminal apex 
dies out gradually while the intensity shifts to the spherical zone resulting 
in the transition of $v_{a}$ and $v_{z}$ from superluminal to subluminal 
values, also seen in Fig.~\ref{FigKaldjoonel3kiirust}.
\end{enumerate}

 Hence, these conclusions are in accordance with the results of the preceding 
Section, despite the fact that the group-velocity expressions used there are in principle 
determined for a quasi-monochromatic pulse. Moreover, as shown in Appendix A, reshaping of the wideband pulse in the course of propagation is in agreement with Eq.~(\ref{BGDelay}) and the curves of group velocity obtained in the preceding section.

Thanks to the symmetry property of the field, $\Psi ( -z ,r , -t) =\Psi ^{ \ast 
}(z ,r ,t)$, and hence $\vert \Psi ( -z ,r , -t)\vert  =\vert \Psi (z ,r 
,t)\vert $, from Fig.~\ref{FigBXevolution} it is easy to comprehend the behavior of the pulse at negative times, i.e., when the pulse converges to the focus.
In the mirrored interpretation the rightmost frame in Fig.~\ref{FigBXevolution}
represents the 
pulse at the earliest time instant   $t = -10$ ps (or its crossing the plane
$z = -3$ mm) and the leftmost frame corresponds now to the latest time instant  
$t =0$ when the pulse has converged to the focus.
Thus, the behavior of the pulse at negative times is the following: initially, 
near  $z = -3$ mm, the BG pulse is given a head start relative to the reference 
(the plane-wave ``signal'' pulse); thereafter, when reaching  $z \approx  -1$ 
mm, due to its subluminality the BG pulse has lost the advantage and the 
signal catches up with it, and at $z \approx  -0.5$ mm, the pulse's peak is 
clearly lagging behind the signal; in the final stage of the focusing the 
situation reverses---due to its superluminal velocity, the BG pulse catches up 
with the signal when they both reach the focal plane $z =0$.
From an experimentalist's point of view more convenient is a scheme where the 
pulse under study starts from a plane $z =z_{in} <0$ together with the plane-wave pulse.
If thereafter the pulse arrival time is measured at point $r_{out}$ in a plane  
$z =z_{out} >0$, the total time $\tau $ of its flight is obviously determined 
by the expression which follows from Eqs.~(\ref{BGDelay}) and (\ref{Va}), viz., 
\begin{align*}\tau  =\tau (\vert z_{in}\vert  ,r_{in}) +\tau (z_{out} 
,r_{out}) = \\
 =\frac{\vert z_{in}\vert }{v_{a}(z_{in} ,r_{in})} 
+\frac{z_{out}}{v_{a}(z_{out} ,r_{out})}\thinspace  ,
\end{align*}
while the reference pulse needs for traveling the same distance a time interval 
$(\vert z_{in}\vert  +z_{out})/c$.
 As a matter of fact, in an experiment, instead of travel times $\tau $,  it is 
convenient to measure delays with respect to the plane wave ``signal'' pulse.
In this case the position $z_{in}$ of the input (start) plane is where the 
delay is initially zero.

It is appropriate to add here that for a BG pulse with fixed geometrical 
parameters, the propagation depth (or temporal interval) where the pulse 
profile transforms from the double-conical (X-like in the meridional section) 
shape to the spherical zone, is nearly independent from the number of cycles in 
the pulse.
What increases with $n$ (while $\omega _{m}$ is kept constant by 
correspondingly adjusting the parameter $t_{0}$) is the duration of the pulse 
only---''thickness'' of its profile in the plots.

In conclusion, despite the fact that the plots in Fig.~\ref{FigBXevolution} have been 
calculated for an 
ultra-wideband pulse, the features of propagation of the BG pulse that they show are 
in accordance with the group velocities defined and studied in Sec. II.

\section{Solving the paradox of spatially averaged group velocity}

The group velocity defined in \cite{Gio}---let us call it the 3D averaged 
velocity---is averaged not only over a distance from $z_{1}$ (particularly, 
$z_{1} =0$) to $z$ [see Eq.~(\ref{Va})], but also over the beam cross section 
from the optical axis to infinity by using the transversal profile of the beam 
as a weighting function.
As a result of such averaging, the effective group velocity is given by an 
expression which is always subluminal and holds for all paraxial 
beams:\begin{equation}v_{3D} =\frac{c}{1 +\frac{\left \langle \mathbf{k}_{ \bot 
}^{2}\right \rangle }{2k^{2}}}\  , \label{V3D}
\end{equation}
where $\left \langle \mathbf{k}_{ \bot }^{2}\right \rangle $ is the dispersion 
(variance) of the transverse wave vector in the beam.
In the case of higher-order Gaussian beams, e.g., Laguerre-Gaussian beams, the 
quantity $\left \langle \mathbf{k}_{ \bot }^{2}\right \rangle $ grows with the 
indices of the beam \cite{Bareza}.

The velocity $v_{3D}$ is not only always less than $c ,$ but also it 
does not depend on the propagation distance.
This is apparently in variation not only with the results of the preceding 
sections, but also with what is well known about group velocities of different 
types of Gaussian, etc.
beams.
For example, for both axicon-generated and circular-grating-generated Bessel 
beams, the transverse (radial) wave vector has a fixed (non-dispersed) value 
$k_{r} =k\sin \theta $, but only the latter beam possesses a subluminal 
velocity in accordance with Eq.~(\ref{V3D}).
For both of these beams, the group velocities do not depend on position and hence 
any averaging over the radial coordinate $r$ must give a value equal to the 
on-axis velocity, which is subluminal in the case of grating-generated Bessel 
beams (or so-called pulsed Bessel beams) and superluminal for axicon-generated 
Bessel beams (or the Bessel-X pulses).
This is easily seen from the following derivation which we will also use later on.

The requisite \textit{harmonic} averaging of  the velocity $v_{a}$ reduces to 
arithmetic averaging of the delay Eq.~(\ref{Delay}) which according to 
\cite{Gio} can be carried out by the following expression for the delay:
\begin{equation}\tau (z) =\frac{ \partial }{c \partial k^{ \prime }}\left
 \{\arg \left [\int \psi ^{ \ast }(z ,r ,k)\thinspace \psi (z ,r ,k^{ \prime 
})\thinspace dS\right ]\right \}_{k^{ \prime } =k}\ \  , \label{Delay3D}
\end{equation}
where $dS =rdr\thinspace d\phi $ is an element of the beam cross section area 
and integration over $\phi $ gives a factor $2\pi $ due to the cylindrical 
symmetry.
The integral over $r$ is proportional to the orthogonality condition between 
Bessel functions and although it diverges if $k^{ \prime } =k$, it is a real 
quantity.
Hence, the $k^{ \prime }$-dependent phase enters into Eq.~(\ref{Delay3D}) only 
from the phase exponent factor of the beams, which is equal to $\exp (iz\sqrt{k^{ 
\prime 2} -k^{2}\sin ^{2}\theta })$ for the pulsed Bessel beam and to $\exp 
(izk^{ \prime }\cos \theta )$ for the Bessel-X pulse.
Taking the derivative $ \partial / \partial k^{ \prime }$ of the phases  and 
using division like in Eq.~(\ref{Va}) we reach a subluminal value $c\cos \theta 
$  of the 3D-averaged group velocity for the pulsed Bessel beam---which reduces 
to Eq.~(\ref{V3D}) in the paraxial limit---and to a superluminal value $c/\cos 
\theta $  for the Bessel-X pulse.
In other words, we reach the well-known results \cite{LW2,PRLmeie,exp2,exp3,meieXfemto,meieOPNis,LiuPulsedBB,DifRefAxiconBB,PorrGaussjaPBB,meieDifAxicon0,meieDifAxicon1,Minupeatykk}.
Note that the transversal averaging has no effect on occasions when the phase 
of the wave function does not depend on the transversal coordinates irrespective 
of the intensity profile of the beam.
This is not the case with the fundamental and higher-order Gaussian beams for 
which the velocity $v_{a}$ becomes subluminal with increasing $r$ even if it 
is superluminal on the optical axis.
Since all Gaussian beams are solutions of the paraxial wave equation and also 
Eq.~(\ref{V3D})  follows from Eq.~(\ref{Delay3D}) in the paraxial limit 
\cite{Gio}, one may suspect that the paraxial approximation is a cause of the 
described discrepancy.

The BG pulse is a convenient object for studying the sources of the discrepancy 
because (i) as we saw in Sec. II, the superluminal values of $v_{a}(z ,r)$ in 
the vicinity of the origin are replaced by subluminal values as $r$ increases, 
and (ii) in the limit $z_{R} \rightarrow \infty $ or $\theta _{0} \rightarrow 0$  
the BG pulse transforms to the Bessel-X pulse.
Unfortunately, for the integrals with the Bessel functions that we encounter in 
Eqs.~(\ref{V3D}) and (\ref{Delay3D}), no closed-form expressions can be found 
in tables of integrals.
Therefore, we carry out the study on a 2D analog of the BG pulse, which, 
however, possesses all the features  essential here.
The wave function of a monochromatic constituent of such a 2D pulse---we name 
it the cosine-Gaussian (CG) pulse---is a solution of the 2D paraxial equation 
and is identical to Eq.~(\ref{BGE}) if there $r$ is replaced by a transversal 
coordinate $x$, the first fractional factor is put under the square root sign, 
and the Bessel functions are replaced by their 2D counterparts---sine and 
cosine function.
Thus, the expression for the delay is similar to Eq.~(\ref{BGDelay}), 
\textit{viz.,}
\begin{multline}\tau (z ,x) =\frac{z}{c} +\frac{1}{c}\frac{z\left (x^{2} 
-\theta ^{2}z_{R}^{2}\right )}{2(z^{2} +z_{R}^{2})} + \\
 +\frac{1}{c}\ensuremath{\operatorname*{Im}}\genfrac{\{}{\}}{}{}{i\theta 
rz_{R}\thinspace \sin \genfrac{(}{)}{}{}{i\theta rz_{R}k}{z +iz_{R}}}{\left (z 
+iz_{R}\right )\thinspace \cos \genfrac{(}{)}{}{}{i\theta rz_{R}k}{z +iz_{R}}} 
, \label{CGDelay}\end{multline}where $\theta $ now is not the axicon angle but 
designates the angle under which two 2D Gaussian beams are inclined with 
respect to the  $\thinspace z$ axis (see Fig. \ref{FigTelgjoonel3kiirust} which
now depicts the beam section in plane $(z ,x)$ at any value of the third 
coordinate $y$).

Harmonic averaging of  the velocity $v_{a}$ is, in other words, a 
cross-section-averaging of the reciprocal velocity which---if normalized with 
respect to $c$---in the plane $z =0$ reads
\begin{multline}\frac{c}{v_{a}\left (0 ,x\right )} =\lim _{z \rightarrow 
0}\frac{c\tau (z ,x)}{z} = \\
 =1 -\frac{\theta ^{2}}{2} +\frac{x^{2}}{2z_{R}^{2}} +\frac{\theta x\tan 
(\theta kx)}{z_{R}} +\frac{\theta ^{2}kx^{2}}{z_{R}\cos ^{2}(\theta kx)} .
\label{1/Va}\end{multline} A limit of Eq.~(\ref{1/Va})
$z_{R} \rightarrow \infty $ results in a superluminal value $v_{a} =c/(1 
-\theta ^{2}/2) \approx c/\cos \theta $, which is simply the group velocity 
of propagation of the interference pattern between two plane waves---the 2D 
counterpart of the Bessel beam.

The averaging can be carried out by the following 
integration\begin{equation}\frac{c}{v_{aa}} =N^{ -1}\int \psi ^{ \ast }(0 ,x 
,k)\frac{c}{v_{a}\left (0 ,x\right )}\psi (0 ,x ,k)\thinspace dx\thinspace  , 
\label{1/Vaa}
\end{equation}where the subscript $aa$ stands for the two-dimensional averaging 
(along the  $z$ and $x$ axes) and $N =\int \psi ^{ \ast }(0 ,x ,k)\psi (0 ,x 
,k)\thinspace dx\thinspace $ is the normalization factor.
The intensity of the beam in the given case reduces to
\begin{equation}\psi ^{ \ast }(0 ,x ,k)\psi (0 ,x ,k) =\cos ^{2}\left (\theta 
kx\right )\thinspace \exp \left ( -\frac{kx^{2}}{z_{R}}\right ) .
\label{weight}
\end{equation}
With the weighting function given by Eq.~(\ref{weight})  the integration 
in Eq.~(\ref{1/Vaa}) can be carried out by each term of  Eq.~(\ref{1/Va}) with 
the help of tables and/or a symbolic calculation software.
The result is the following:\begin{equation}\frac{c}{v_{aa}} =1 +\frac{\theta 
_{0}^{2}}{8} +\frac{\theta ^{2}\exp \genfrac{(}{)}{}{}{\theta ^{2}}{\theta 
_{0}^{2}}}{4\cosh \genfrac{(}{)}{}{}{\theta ^{2}}{\theta _{0}^{2}}}\thinspace  
, \label{1/VaaFIN}
\end{equation}
where we have replaced  $k$ and $z_{R}$ with a more transparent parameter---the 
Gaussian beam divergence angle $\theta _{0}$ according to the known relation 
$kz_{R} =2/\theta _{0}^{2}$.
It is immediately evident from Eq.~(\ref{1/VaaFIN}) that the fully averaged 
group velocity of the CG pulse is \textit{subluminal}.
The second term is the contribution to the subluminality that comes from the 
divergence of the wave vectors of the plane-wave constituents of the Gaussian 
beam (in the case of a 3D Gaussian beam, this contribution is $\theta _{0}^{2}/4$  
due to the two-times-larger number of the transversal dimensions).
Even if we go to the limit of two plane waves in Eq.~(\ref{1/VaaFIN}) by 
letting $\theta _{0} \rightarrow 0$, the result $1 +\theta ^{2}/2$  
nevertheless corresponds to subluminal velocity $v_{aa} =c/(1 +\theta ^{2}/2) 
\approx c\cos \theta $.
The same results follow from the averaging procedures with the help of 2D 
versions of Eq.~(\ref{Delay3D}) and (\ref{V3D}).

If we \textit{first} go to the limit of two plane waves---as we did 
above after  Eq.~(\ref{1/Va})---and \textit{thereafter} carry out the averaging 
by the integration, we obviously get the \textit{superluminal} velocity $c/\cos 
\theta $.
The mathematical reason for such contradiction is that taking a limit and an 
improper integral need not to be interchangeable operations.
The physical reason is that the Gaussian aperturing introduces a dependence of 
the group velocity $v_{a}$ on a transverse coordinate and $v_{a}$ generally 
decreases with the increase of the distance from the optical axis as we saw 
above.

These conclusions can be readily generalized to the 3D case: despite the fact that in the 
focus the BG pulse is superluminal, its fully averaged group velocity is 
subluminal.
In contrast, the group velocity of the Bessel-X pulse is superluminal 
irrespective of how large the averaging area is.
In this regard, it is interesting to evaluate the maximum radius $r_{sl}$ of 
such disk of averaging, at which the averaged velocity of the BG is still 
superluminal.
For the parameters used in Sec. II (indicated in Fig. \ref{FigTelgjoonel3kiirust})
with the help of numerical integration we obtained $r_{sl} \approx 0.68w_{0} 
=17.8 \mbox{{\textmu}m}$.
For experimental observation of the superluminality of the BG pulse it means 
that the diameter of the acceptance area of the probing or detecting device must 
be smaller than $ \sim 34\mbox{{\textmu}m}$.
This is not a problem when the tip of an optical fiber is used as a probe, in 
which case an order of magnitude better resolution can be achieved 
\cite{meieXfemto,meieDifAxicon1}.
The circle with $r_{sl} \approx 0.68w_{0}$ in the focal plane cuts the beam at 
the level 10\% of its peak intensity, i.e., it embraces the pulse central peak 
almost completely. Nevertheless, as shown in Appendix B, the 
contribution to the averaged velocity, which originates from all the infinite 
off-axis area with $r >r_{sl}$, outweighs the superluminal contribution 
resulting in a subluminal value of the averaged velocity $v_{3D}$, in accordance 
with Eq.~(\ref{V3D}).

\textit{The presented analysis does not say that the three-dimensionally averaged group 
velocity  is somehow more proper than the mean velocity} $v_{a}$ \textit{defined by} 
Eq.~(\ref{Va}).
Both are directly related to experimentally measurable propagation times or 
delays, but the first quantity is applicable if the pulse is recorded without 
any spatial resolution in a  transverse plane and the second one if the 
dependence of the delay not only on the propagation depth but also on the  
transverse coordinates is measured.

\section{Conclusions}

Our general conclusion is that an answer to the question of which version of group 
velocity the pulse time of flight between two planes (which are perpendicular 
to the pulse propagation axis) is directly related to depends on how the pulse 
is registered.

If its intensity is measured not only with temporal but also with spatial 
resolution so that for the peak or any other local feature at point 
$\mathbf{R}$ of the pulse profile the travel time $\tau (\mathbf{R})$ between 
the two planes is recorded, then the travel distance divided by $\tau 
(\mathbf{R})$ results in the group velocity $v_{a}\left (\mathbf{R}\right )$ of 
that feature in the direction of the optical axis averaged over the travel 
distance.
This version of averaged group velocity may be subluminal, luminal or 
superluminal.
By varying the position of the output plane near the beam focus, a detailed 
picture of the behavior of group velocity in the focal region can be obtained.
In the far field the Bessel-Gauss\ pulse and, generally, pulses of 
Laguerre-Gauss and other hollow beams propagate like an expanding spherical 
zone and therefore their Born-Wolf group velocity \cite{BornWolf} equals 
$c$, i.e., is luminal.
But measuring the time of flight of the pulse peak in a usual optical scheme 
with the output (detecting) plane in the far field exhibits a subluminal 
propagation, no matter how the group velocities behave in the focal region of these beams.
The reason is that outside the Rayleigh range, the velocity $v_{a}\left 
(\mathbf{R}\right )$ as well as the projection of the Born-Wolf group-velocity 
vector onto the propagation axis reach their common subluminal constant value 
determined by beam divergence.
If the pulse is registered as a whole in the output (detecting) plane without  
spatial resolution, then the travel distance divided by the travel time $\tau $ 
of the pulse results in the three-dimensionally averaged group velocity 
$v_{3D}$ introduced in \cite{Gio,Bareza}.
For paraxial pulsed beams, this velocity (i) does not depend on the position of the planes and (ii) is always subluminal.
These properties are not in contradiction with studies of  superluminally propagating Bessel-X-type nondiffracting pulses for two reasons:
\begin{enumerate}
\item Their superluminality has been examined with spatial resolution in their cross-sectional plane, and
\item any physically realizable Bessel-X-type pulse has a finite aperture like the 
Bessel-Gauss pulse has the Gaussian aperture and therefore the cross-sectional 
averaging of the group velocity results in a subluminal value as was shown here for the Bessel-Gauss pulse.
\end{enumerate}

The results obtained here also hold in the case of single-photon pulses as far as spatio-temporal dependencies in photon wave functions are the same as in classical wave packets.


\begin{acknowledgments}
This research has been supported by the Estonian Research Council through Grant No. PUT369. 
 The author thanks Ioannis Besieris, Daniele Faccio, John Lekner, Miguel 
Porras, and V{\'a}clav Poto\v{c}ek for fruitful discussions, as well as Heli 
Lukner, Peeter Piksarv, and Andreas Valdmann for their valuable comments on the 
manuscript.
\end{acknowledgments}

\appendix
\section{SUBTLETIES OF PROPAGATION OF THE BG PULSE}
A closer look at the first three frames of Fig. 6 reveals that the pulse's peak slightly 
lags behind the X-shape symmetry plane given by the zeros of the Bessel 
profile, but nevertheless it moves superluminally.
At the same time curve 3 in Fig. \ref{FigTelgjoonel3kiirust} tells us that the bottom of the pulse's peak should propagate up to $z 
\lesssim 3/4\ z_{R}$ subluminally and therefore its shape in the frames 2-6 in 
Fig. 6 should gradually become distorted from an oval to a horseshoe shape like ''$ 
\supset $.''
 A closer study of this discrepancy reveals that it is caused solely by the 
3.6-fold difference of the axicon angles that we have chosen in the preceding and 
the current section for clarity of the plots.
Indeed, with the value $\theta  =2^{\circ}$, curve 3
in Fig. \ref{FigTelgjoonel3kiirust} obtains a shape corresponding to 
superluminal values of the velocity 
$v_{a\text{}}$ over the whole range of distances $z$.
Vice versa, plotting with the value $\theta  =0.55^{\circ}$, the first frames of Fig.
6 with high resolution in the region of the central peak, the brightness 
distribution gradually changes indeed to a horseshoe shape.

There is another subtlety seen in the first frame, viz., at larger radial 
distances than the Bessel profile zeros there are two maxima (the ``branches'' 
of X shape), one of which moves through the plane $z =0$ before the instant $t 
=0$ and the other later.
At the same time Eq.~(\ref{BGDelay}) turns to zero at $z =0$, irrespective of 
the radial coordinate, which means that the intensity at these radial distances 
should reach its peak values at  $t =0$, i.e., in the middle between the 
branches where actually the frame shows the lowest intensities.
A contradiction? No, because, first of all, in the case of the 
quasimonochromatic limit assumed by the notion of group velocity and, 
consequently, by Eq.~(\ref{BGDelay}), the frame would be fulfilled with a 
Bessel beam whose temporal profile at all radial distances would peak at $t 
=0$, $z =0$, and the ``X branching'' (formation of the double-conical profile) 
would occur at very large radial distances $r_{X} \sim cT/2\theta $, where $T$  
is the duration of the pulse.
Parenthetically, $r_{X}$ would encircle as many Bessel profile zeros, as many 
cycles the pulse contains.
Moreover, there is no contradiction with Eq.~(\ref{BGDelay}) even in the case 
of the branches of the ultrawideband pulse in Fig.~\ref{FigBXevolution}.
Namely, if one plots from  Eq.~(\ref{BGDelay}) the temporal shift  $z/c -\tau $ 
(for $r =r_{X}$ evaluated from the pulse parameters of Fig.~\ref{FigBXevolution})
as a function of 
$z$, one gets a curve which with increasing $z$ immediately after the origin $z 
= +0$  jumps to negative values and then in the vicinity of the point $z \approx 
0.5z_{R}$ turns to positive values.
This is in accordance with the behavior we see in Fig.~\ref{FigBXevolution} also.
Indeed, when the pulse center has passed the plane $z =0$, the branches are 
no longer of equal intensity, the field modulus at $r =r_{X}$ has it maximum on 
the rear cone, and, since this cone moves behind the apex, the temporal shift is 
negative.
With increasing $z$, however, since the whole double-conical part of the pulse 
profile moves superluminally, the beginning point of the rear (left) branch 
at $r =r_{X}$ gradually recovers from its initial lag and for it the shift  
$z/c -\tau $ becomes positive (note that here we  are looking upon propagation 
of intensity or modulus maxima at a fixed radial distance from the $z$ axis and 
not the propagation of the absolute intensity maximum of the pulse).

\section{SUPER- AND SUBLUMINAL CONTRIBUTIONS TO THE 3D-AVERAGED VELOCITY}
Here we evaluate how much power (intensity 
integrated over area)  the disk with radius $r_{sl}$ transmits in comparison to the 
total power (intensity integrated over whole plane $z =0$) at the instant $t 
=0.$  Concerning the evaluation of the improper integral for the latter quantity, a 
numerical calculation would have been dubious, because the integrand contains the 
infinitely many times oscillating Bessel function; see Eq.~(\ref{BGE}).
Fortunately, we found a table integral given by Eq.~(6.633.4) in Ref.
\cite{Gr-Rtabel} which readily gave an exact result:
\begin{multline}\int _{0}^{\infty }\psi ^{ \ast }(0 ,r ,k)\psi (0 ,r 
,k)r\thinspace dr = \\
 =\int _{0}^{\infty }e^{ -\frac{kr^{2}}{z_{R}}}\thinspace J_{0}^{2}\left 
(\theta kr\right )r\thinspace dr = \label{totalpower} \\
 =k^{ -1}z_{R}I_{0}\left (\theta ^{2}kz_{R}\right )\exp \left ( -\theta 
^{2}kz_{R}\right ) ,\end{multline}where $I_{0}( .)$ is the modified Bessel 
function of the zeroth order.
With the help of Eq.~(\ref{totalpower}) we found that in the case of the BG 
pulse, with the given beam parameters, as much as about 70\% of the total power 
is embraced within the superluminality circle  of radius $r_{sl}$. We conclude that 
the subluminal contribution to the averaged velocity, which originates from the 30\% of the total power flowing in the off-axis area with $r >r_{sl}$, outweighs the superluminal contribution resulting in a subluminal value of the averaged velocity $v_{3D}$.

\end{document}